\documentclass{ws-mpla}
\usepackage[super]{cite}
\usepackage{graphicx}

\begin{document}

\markboth{A. Tartaglia et al.}
{Hunting for a new test of general relativity}

\catchline{}{}{}{}{}

\title{LOOKING FOR A NEW TEST OF GENERAL RELATIVITY\\
IN THE SOLAR SYSTEM}

\author{ANGELO TARTAGLIA}

\address{Politecnico di Torino, Corso Duca degli Abruzzi 24, 10129 Torino, Italy\\
INdAM, Citt\`a Universitaria, Piazzale Aldo Moro 5, 00185 Roma, Italy\\
angelo.tartaglia@polito.it}

\author{GIAMPIERO ESPOSITO}

\address{Istituto Nazionale di Fisica Nucleare, Sezione di Napoli,\\
Complesso Universitario di Monte S. Angelo,\\
Via Cintia Edificio 6, 80126 Napoli, Italy\\
gesposit@na.infn.it}

\author{EMMANUELE BATTISTA}

\address{Istituto Nazionale di Fisica Nucleare, Sezione di Napoli,\\
Complesso Universitario di Monte S. Angelo,\\
Via Cintia Edificio 6, 80126 Napoli, Italy\\
ebattista@na.infn.it}

\author{SIMONE DELL'AGNELLO}

\address{Istituto Nazionale di Fisica Nucleare,\\
Laboratori Nazionali di Frascati, 00044 Frascati, Italy\\
simone.dellagnello@lnf.infn.it}

\author{BIN WANG}

\address{IFSA Collaborative Innovation Center, School of Physics and Astronomy,\\
Shanghai Jiao Tong University, Shanghai 200240, China\\
Center for Gravitation and Cosmology, College of Physical Science and Technology,\\
Yangzhou University, Yangzhou 225009, China\\
${\rm wang}_{-}{\rm b}$@sjtu.edu.cn}

\maketitle

\pub{Received (Day Month Year)}{Revised (Day Month Year)}

\begin{abstract}
This paper discusses three matter-of-principle methods 
for measuring the general relativity correction to the
Newtonian values of the position of collinear Lagrangian points $L_{1}$ and $L_{2}$ of
the Sun-Earth-satellite system. All approaches are based on time measurements. 
The first approach exploits a pulsar emitting signals and
two receiving antennas located at $L_{1}$ and $L_{2}$, respectively. The second method
is based on a relativistic positioning system based on the Lagrangian points themselves.
These first two methods depend crucially on the synchronization of clocks at $L_{1}$ and
$L_{2}$. The third method combines a pulsar and an artificial emitter at the stable points
$L_{4}$ or $L_{5}$ forming a basis for the positioning of the collinear points $L_{1}$ and $L_{2}$.
Further possibilities are mentioned and the feasibility of the measurements is considered.

\keywords{$3$-body problem; Lagrangian points; general relativity.}
\end{abstract}

\ccode{PACS Nos.: 04.20Cv, 95.10.Ce}

\section{Introduction}	

From the age of Laplace and Poincar\'e \cite{Laplace,Tisserand,Poincare}
until recent times \cite{KP1,KP2}, celestial mechanics has played
the crucial role of providing a testbed for the theories of gravitation that mankind has been
able to develop. When Poincar\'e discovered chaos in his investigation of $3$-body dynamics in
Newtonian gravity, he stressed this peculiar role of celestial mechanics at the beginning of
his monumental treatise on the new methods of celestial mechanics \cite{Poincare}.
A quarter of a century later, after Einstein arrived at his geometric view of gravitation, the
precession of Mercury's perihelion, observed by the astronomers, turned out to be in complete
agreement with the calculation based upon the
geodesic motion of planets in Einstein's theory \cite{Choquet},
and since then his general relativity has passed many
observational tests \cite{Will}, including the recent discovery of gravitational waves \cite{Abbott}.
However, in the light of the accelerated expansion of the universe, which seems to cast doubts on the
attractive nature of gravity on all scales, and bearing in mind the attempt to
question the existence of dark matter by appealing to modified gravitational
Lagrangians \cite{Capozziello2011} (cf. Ref. \cite{Wang2017}),
satellite and planetary motions in the solar system are still receiving careful consideration,
with the hope of being able to discriminate between Einstein's theory and extended gravity
theories \cite{Martini,Ciocci}.

Over the last few years, some of us have looked, in particular, at the tiny corrections to the
location of Lagrangian points, both stable and unstable, in the Earth-Moon-satellite
\cite{Battista2014a,Battista2014b,Battista2015a,Battista2015b} and
Sun-Earth-satellite \cite{Battista2017a} $3$-body systems.
In the present letter we are concerned with the purely
classical corrections that Einstein's theory predicts for the location of collinear Lagrangian
points, with respect to Newtonian gravity. We had originally looked at the Earth-Moon-satellite
system, but the corresponding corrections are a few millimeters only. This theoretical
value is conceptually interesting but is still too small, despite the extraordinary progress
made by laser ranging techniques, that have reached the centimeter accuracy by now. Moreover,
for the location of collinear Lagrangian points $L_{1}$ and $L_{2}$
of the Sun-Earth-satellite system, we have recently found
that Einstein's theory corrects Newtonian gravity by $5$ meters and $-4.8$ meters,
respectively \cite{Battista2017a}. These corrections are apparently big enough to allow for experimental verification.

It is now our aim to discuss the possibility of measuring such an effect, which seems
to be a good example of Galilean physics within mankind's reach.

\section{Approach (a): pulsars}

Let us start from the work in Ref. \cite{Battista2017a}, and in particular from Table 3
therein, which displays an important general relativistic correction to the location
of collinear Lagrangian points $L_{1}$ and $L_{2}$ for a $3$-body system consisting
of Sun, Earth and a satellite, with respect to the Newtonian gravity. The amount of 
the correction has been mentioned in the introduction; considering the effect on both 
$L_1$ and $L_2$, the two points turn out to be $9.8$ m closer to one another. We may 
think of detecting such change of distance between the two collinear points, at least 
in principle, by measuring the time of flight difference of an electromagnetic 
signal between $L_1$ and $L_2$.

In the first approach that we present, our idea it to use signals emitted from one 
or more pulsars (for redundancy). Once we have chosen a specific pulsar, suppose that 
two receiving antennas (radiotelescopes) are available, one at $L_{1}$ and the other
at $L_{2}$. The wavefronts of the pulses emitted from the pulsar first run over $L_{2}$ then
$L_{1}$ (or in the reverse order). The difference in the arrival times depends on the distance
between the two points and on the angle $\alpha$ between the $L_{1}$-$L_{2}$ axis and the
direction of the pulsar (assumed to be at infinity). The setting is shown in Fig. 1 below.

\begin{figure}[ph]
\centerline{\includegraphics[width=4.0in]{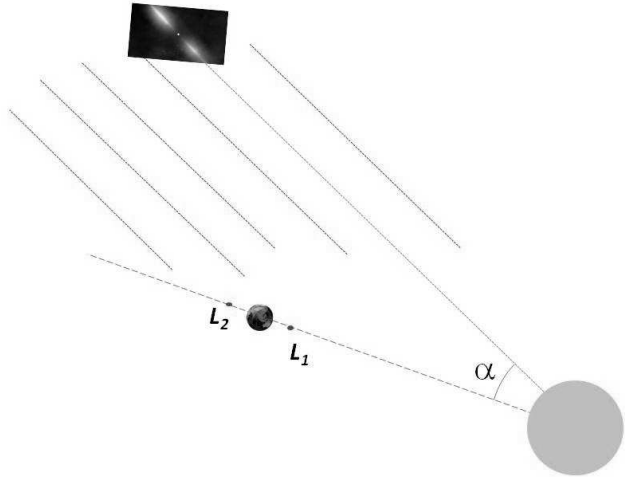}}
\vspace*{8pt}
\caption{On top we represent the pulsar that emits signals, while $L_{1}$ and $L_{2}$
are two collinear Lagrangian points of the Sun-Earth-planetoid system, and $\alpha$
is the angle between the line joining such points and the direction of the pulsar.}
\end{figure}

On denoting by $a$ the distance between $L_{1}$ and $L_{2}$, the arrival times difference
$\tau$ is
\begin{equation}
\tau={a \over c}\cos(\alpha).
\label{(1)}
\end{equation}
The approximate value of $a$ is
\begin{equation}
a \approx 3 \times 10^{9} \; {\rm m},
\label{(2)}
\end{equation}
and hence
\begin{equation}
\tau \leq 10 \; {\rm s}.
\label{(3)}
\end{equation}
As we have already seen, according to Ref. \cite{Battista2017a}, the general relativity 
correction to $a$ amounts to a variation $\delta a=9.8$ 
meters, which implies a relative variation
\begin{equation}
{\delta a \over a} \approx 3 \times 10^{-9}.
\label{(4)}
\end{equation}
The resulting expected change in the times of arrival difference would be
\begin{equation}
\delta \tau \leq 3 \times 10^{-8} \; {\rm s},
\label{(5)}
\end{equation}
which means that the accuracy in the measurement of time should of course be better
than that. Moreover, it implies that the Newtonian positions of collinear points
$L_{1}$ and $L_{2}$ should be known with accuracies better than at least $1$ meter.

As far as the choice of the pulsar is concerned, one should look for a sufficiently bright
one (although it is known that they are, in general, very faint objects). The available
periods range from a few $ms$ to a few $s$, and the relative stability of the emission rate
(rotation of the star) may be $1$ part in $10^{12}$ per cycle. Single pulses are usually not identical
to one another, however, for the application we are considering here, what really matters
is the possibility to recognize one and the same pulse after it has passed through the
Lagrangian points $L_{1}$ and $L_{2}$.

In practice, one should record a sequence of pulses both at $L_{1}$ and at $L_{2}$, then the
two series should be confronted in the same time reference and shifted until 
they coincide. The duration of the shift would measure
the difference in the arrival times; it should be measured with an accuracy better at least
than $10^{-8}$s, in order to make the comparison between the Newtonian and the general
relativity results possible.

Of course, the measurement should be performed by considering a variety of sources
with different positions in the sky and different periods. A non-trivial basic problem is
that one should previously synchronize the clocks at $L_{1}$ and $L_{2}$ with an accuracy
better than $1$ ns.

\section{Approach (b): a relativistic positioning system based upon the Lagrangian points
themselves}

The Lagrangian points of the Sun-Earth-satellite $3$-body system are fit to be the material
bases for a Relativistic Positioning System \cite{Tartaglia2011,Tartaglia2017}.
However in this case the situation, by virtue of peculiar symmetries, looks simpler.
All Lagrangian points lie in the same plane, and the two that are more interesting for us
are located along the Sun-Earth line; in practice, the problem is therefore one-dimensional.

Suppose then to locate beacons at the stable points $L_{4}$ and $L_{5}$ emitting periodic
signals at a stable frequency; they would act with respect to $L_{1}$ and $L_{2}$ more or
less as artificial pulsars, with much stronger signals.

Apparently, only one emitter would be enough (either at $L_{4}$ or at $L_{5}$). The emission
sequence from, say, the stable point $L_{4}$, since the emitter is at rest with respect to
$L_{1}$ and $L_{2}$, coincides with the sequence of arrivals at $L_{1}$; the unstable
point $L_{2}$ will receive the same sequence, with some delay depending on the greater distance
of the receiver from the emitters. The procedure is then the same as for the pulsars in
the previous section, and also for the accuracies involved the same considerations can be made.
The difficulty that is challenging us is also the same: the clocks at collinear points
$L_{1}$ and $L_{2}$ must be synchronous.

\section{Approach (c): stable Lagrangian points and a pulsar}

In order to avoid the need for synchronization of the clocks, the full Relativistic
Positioning System should be used. In practice, at least one source of pulses out of the
ecliptic plane is required, and this can be a pulsar. The pulsar and an artificial
emitter at $L_{4}$ or $L_{5}$ can provide the positioning of unstable Lagrangian points
$L_{1}$ and $L_{2}$ both in space and time, i.e. to give the relative distance between them.

As for the practical aspects of measurements as the ones we have suggested here, we must
add that, of course, we would not have spacecrafts permanently located at $L_{1}$ and
$L_{2}$. Both positions are indeed unstable, so that the receivers therein would move
around the corresponding Lagrangian point along Lissajous orbits. Consequently the
measurement should be repeated several times. The reference would then be to the average
position over time, the center of the instantaneous positions being coincident with
the Lagrangian point.

\section{Closed contours}
\begin{figure}[ph]
\centerline{\includegraphics[width=4.0in]{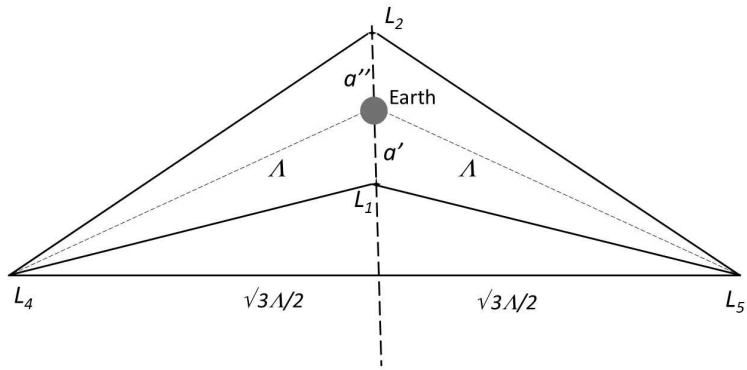}}
\vspace*{8pt}
\caption{The scheme (not on scale) represents the possible closed paths including the 
Lagrangian points to be used to send electromagnetic signals and measure the 
times of flight along closed paths.}
\label{fig2}
\end{figure}

The difficulty of synchronizing clocks located million kilometers apart from one another 
is not easy to surmount. A way to solve the problem would be to use just one clock. 
This would be possible if the electromagnetic signals move along a closed path in space, 
starting from and arriving to the same position. Consider for instance a path 
$L_1-L_4-L_2-L_5-L_1$; its length (see fig. \ref{fig2}) would be
\begin{equation}
l=2 \left(\sqrt{\Lambda^2+\Lambda a''+a''^2}+\sqrt{\Lambda^2-\Lambda a'+a'^2} \right).
\label{(6)}
\end{equation}
With our notation, $\Lambda$ is the distance of $L_4$ or $L_5$ from the Earth and corresponds 
to $1$ AU; $a'$ is the distance of $L_1$ from the Earth and $a''$ is the 
distance of $L_2$ from the Earth; of course $a'+a''=a$.

On taking into account that $a'\simeq a''\sim \frac{\Lambda}{10}$, the total time of flight 
of light moving along the path would be approximately $2000$ s. A change $\delta a$ 
of the distance $a$ would produce a time of flight change
\begin{equation}
\delta \tau\simeq 5\times 10^{-9} \; {\rm s}.
\label{(7)}
\end{equation}

A more favorable configuration would be obtained by using the triangle $L_2-L_5-L_4-L_2$ 
(see fig. \ref{fig2}). In that case the total path would be 
$l=2\sqrt{\Lambda^2+a''\Lambda+a''^2}+\sqrt{3} \Lambda$; the corresponding time of flight 
is approximately $1920$ s, and the change in the time of flight induced 
by a change $\delta a'' = 5$ m would be roughly 

\begin{equation}
\delta \tau\simeq 2\times 10^{-8} \; {\rm s}.
\label{(8)}
\end{equation}

\section{Discussion}

The models we have considered have a matter-of-principle nature. A number of practical 
problems are however present. For example, as shown in detail
by Brumberg, by virtue of the emission of gravitational radiation, the libration points
$L_{1},L_{2},L_{3},L_{4},L_{5}$ of Newtonian theory are, strictly, quasi-libration
points.\cite{Brumberg} Moreover, non-gravitational perturbations must also be accounted for; 
for instance, radiation pressure affects the motion of a
spacecraft,\cite{RP1,RP2,RP3,RP4,RP5,RP6} and hence a proposal to a space agency should
consider carefully also the shape of the desired satellite, especially in the case 
that extended antennas should be deployed.

Leaving such complications aside, one can say what follows.
So far, we have described our original ideas on how in principle the correction in
the position of the Sun-Earth collinear Lagrange points, induced by General
Relativity effects, could be measured. It is not the purpose of the present paper to enter a 
detailed technical analysis, but some comments on the practical feasibility 
and the problems to tackle are in order.

Let us start with the proposal to use pulsars, approach a of Sect. 2. As
mentioned therein, a practical problem is the size of the required antenna, by virtue
of the weakness of signals. It is worth mentioning a solution that would
be viable: using X-ray pulsars. Such sources are much less numerous than the
more common radio-pulsars, but the antennas they need have much smaller size
than radiotelescopes. The use of X-ray pulsars is indeed under study and
development by NASA for space navigation purposes (XNAV system); see for
instance Ref. \cite{NASA}

The other major delicate issue, besides the need for synchronization of the
clocks, is the actual position of the spacecraft that should mark $L_{1}$
and $L_{2}$. Of course, they would move around the corresponding Lagrangian points,
rather than stably stop there. As we said before, the position at collinear points is
unstable and a station therein would move around the Lagrangian
point, tracing planar Lissajous figures or halo orbits, with a period of
approximately six months. The receiver would then slowly drift away. The
whole issue is discussed in Sect. 6 of Ref. \cite{Tartaglia2017}
Upon considering our approach (c), also
$L_{4}$ or $L_{5}$ come into play
and, of course, also for them the problem would arise
of what is the actual position of the spacecraft while performing the
measurement. Now the orbits are stable quasi-periodic, with a period of
order one year.

Last but not least, the accuracy with which the Newtonian position of $L_1$ and $L_2$ 
can be calculated depends on the accuracy on the value of the masses of the Earth and 
the Sun, and on the distance between the two bodies. At the moment the resulting 
uncertainty in the position of the two $L$-points seems to be bigger than the correction 
due to General Relativity. Furthermore the orbit of the earth is not exactly circular, 
which means that the position of both $L_1$ and $L_2$ periodically changes during the year.
 
The above remarks tell us that we should in any case think of a measurement strategy
where we look for the central values in a cloud of results
generated by the instantaneous positions of the space stations 
(and even of the $L$-points). Taking into account
the slowness of the movements about the Lagrangian points, the data acquisition
should last quite a few years.

\section*{Acknowledgments}

E. Battista and G. Esposito are grateful to the Dipartimento di Fisica ``Ettore Pancini''
of Federico II University for hospitality and support.

\end{document}